\documentclass{IOS-Book-Article}

\usepackage{mathtools, mathptmx, amsmath, amsthm, amssymb}
\usepackage{graphicx, subcaption}
\usepackage{listings}
\usepackage{caption}
\usepackage{multirow}
\usepackage{color}
\usepackage{cite, hyperref}
\usepackage{siunitx}
\usepackage{xspace}
\usepackage{makecell}

\definecolor{gray}{gray}{0.3}
\definecolor{graybox}{gray}{0.8}

\newcommand{\swift}{{\sc Swift}\xspace}
\newcommand{\gadget}{{\sc Gadget-2}\xspace}

\def\hb{\hbox to 10.7 cm{}}

\begin{document}

%%%%%%%%%%%%%%%%%%%%%%%%%%%%%%%%%%%%%%%%%%%%%%%%%%%%%%%%%%%%%%%%%%%%%%%%%%%%%%%%
%  Set options for the listings package                                        %
%%%%%%%%%%%%%%%%%%%%%%%%%%%%%%%%%%%%%%%%%%%%%%%%%%%%%%%%%%%%%%%%%%%%%%%%%%%%%%%%
\lstset{%
    language=c,
    basicstyle=\scriptsize\ttfamily,
    keywordstyle=\bfseries,
    commentstyle=\color{gray},
    rulecolor=\color{black},
    framerule=0.6pt,
    numbers=left,
    numberstyle=\tiny,
    escapeinside={@}{@},
    captionpos=b
  }

\pagestyle{headings}
\def\thepage{}

\begin{frontmatter}              % The preamble begins here.

%%%%%%%%%%%%%%%%%%%%%%%%%%%%%%%%%%%%%%%%%%%%%%%%%%%%%%%%%%%%%%%%%%%%%%%%%%%%%%%%
%  Title, author and affiliations                                              %
%%%%%%%%%%%%%%%%%%%%%%%%%%%%%%%%%%%%%%%%%%%%%%%%%%%%%%%%%%%%%%%%%%%%%%%%%%%%%%%%
\title{An Efficient SIMD Implementation of Pseudo-Verlet Lists for Neighbour Interactions in Particle-Based Codes}

\markboth{}{September 2017\hb}

\author[A]{\fnms{James S.} \snm{Willis}%
\thanks{Corresponding Author; E-mail:\url{james.s.willis@durham.ac.uk.}}},
\author[A]{\fnms{Matthieu} \snm{Schaller}},
\author[B,C]{\fnms{Pedro} \snm{Gonnet}}
\author[A]{\fnms{Richard G.} \snm{Bower}}
and
\author[A]{\fnms{Peter W.} \snm{Draper}},

\runningauthor{Institute for Computational
Cosmology (ICC),
Department of Physics,
Durham University,
Durham DH1 3LE, UK}
\address[A]{Institute for Computational
Cosmology (ICC),
Department of Physics,
Durham University,
Durham DH1 3LE, UK}
\address[B]{School of Engineering and
Computing Sciences,
Durham University,
Durham DH1 3LE, UK}
\address[C]{Google Switzerland GmbH,
8002 Z{\"u}rich, Switzerland}

%%%%%%%%%%%%%%%%%%%%%%%%%%%%%%%%%%%%%%%%%%%%%%%%%%%%%%%%%%%%%%%%%%%%%%%%%%%%%%%%
%  Abstract                                                                    %
%%%%%%%%%%%%%%%%%%%%%%%%%%%%%%%%%%%%%%%%%%%%%%%%%%%%%%%%%%%%%%%%%%%%%%%%%%%%%%%%
\begin{abstract}
In particle-based simulations, neighbour finding (i.e finding pairs of particles to interact within a given range) is the most time consuming part of the computation. One of the best such algorithms, which can be used for both Molecular Dynamics (MD) and Smoothed Particle Hydrodynamics (SPH) simulations, is the pseudo-Verlet list algorithm. This algorithm, however, does not vectorise trivially, and hence makes it difficult to exploit SIMD-parallel architectures. In this paper, we present several novel modifications as well as a vectorisation strategy for the algorithm which lead to overall speed-ups over the scalar version of the algorithm of 2.24x for the AVX instruction set (SIMD width of 8), 2.43x for AVX2, and 4.07x for AVX-512 (SIMD width of 16).
\end{abstract}

%%%%%%%%%%%%%%%%%%%%%%%%%%%%%%%%%%%%%%%%%%%%%%%%%%%%%%%%%%%%%%%%%%%%%%%%%%%%%%%%
%  Metadata                                                                    %
%%%%%%%%%%%%%%%%%%%%%%%%%%%%%%%%%%%%%%%%%%%%%%%%%%%%%%%%%%%%%%%%%%%%%%%%%%%%%%%%
\end{frontmatter}
\markboth{J.~Willis, M.~Schaller, P.~Gonnet, R. Bower~\&~P.~Draper 2017 \hb}{SIMD pseudo-Verlet list for particle-based codes\hb}

%%%%%%%%%%%%%%%%%%%%%%%%%%%%%%%%%%%%%%%%%%%%%%%%%%%%%%%%%%%%%%%%%%%%%%%%%%%%%%%%
%  Introduction                                                                %
%%%%%%%%%%%%%%%%%%%%%%%%%%%%%%%%%%%%%%%%%%%%%%%%%%%%%%%%%%%%%%%%%%%%%%%%%%%%%%%%
\section{Introduction}
\label{ref:introduction}

\begin{figure}
  \centering
  \includegraphics[width=0.6\textwidth]{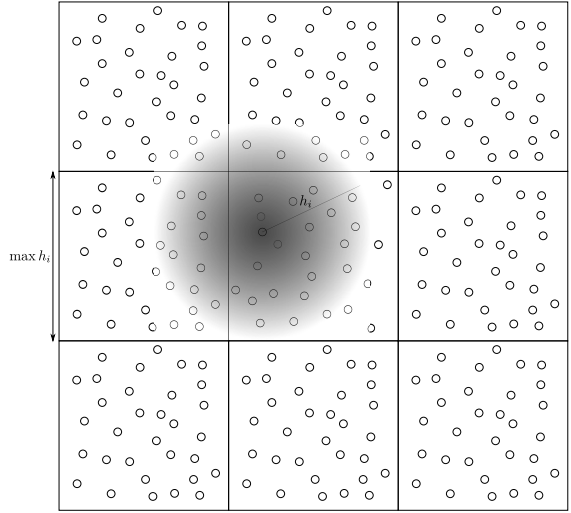}
  \caption{Computing the interactions of a set of $N$ particles using Smoothed Particle Hydrodynamics. Each particle finds its neighbours by computing $N - 1$ pairwise distances and checking whether they fall within a cut-off radius, $h_i$. The particle's density is then found by computing a weighted sum over all its neighbours, where the weight decreases with distance from the particle up to $h_i$.}
   \label{fig:sph}
\end{figure}

Particle-based simulations are widely used in many research fields, e.g. chemistry, physics, biology. The under-lying algorithm evolves a system of particles via a set of pairwise interactions. Short-range interactions are evaluated by computing pairwise distances between particles and checking that they lie within a cut-off radius, $h$, of each other (see Figure~\ref{fig:sph}).

Computing the short-range pairwise distances between particles takes up a large fraction of the CPU time for these simulations. For $N$ particles, a naive implementation involves $\mathcal{O}(N^2)$ operations, which can be expensive to compute. The number of distance calculations can be significantly reduced by using a cell list \cite{ref:Cell_list}, that decomposes the domain into cells of edge length $C_l \geqslant h$ such that each particle can find its neighbours by searching only in the local or neighbouring cells.

However, cell lists are still far from optimal, as it can be shown that for a uniform distribution of particles in two neighbouring cells (sharing a face) only $\leqslant$16\% of particle pairs will meet the criteria $r^2 \leqslant h^2$, where $r$ is the separation between particles \cite{ref:PG_cell_orientations}. This leads to $\geqslant$84\% unnecessary distance calculations between pairs of cells (see Figure~\ref{fig:sph} for a 2D example). This fraction can be reduced using the pseudo-Verlet list algorithm\cite{ref:PG_Pseudo_Verlet_list}. But the pseudo-Verlet list algorithm, due to its branching and inherently inefficient memory access patterns, does not lend itself to automatic SIMD-vectorisation, which is crucial for obtaining the best possible performance on modern CPU architectures. We therefore, in this paper, present a SIMD vectorisation strategy for the pseudo-Verlet list that addresses these issues.

This optimisation strategy can be applied to Molecular Dynamics (MD) and Smoothed Particle Hydrodynamics (SPH) codes widely used in chemistry and physics. We have chosen to implement our SIMD strategy in \swift\footnote{\url{www.swiftsim.com}}, a cosmological simulation code \cite{ref:SWIFT_SIAM, ref:SWIFT_PASC} that solves the equations of hydrodynamics using SPH (See \cite{ref:SPH_review} for a review).

%%%%%%%%%%%%%%%%%%%%%%%%%%%%%%%%%%%%%%%%%%%%%%%%%%%%%%%%%%%%%%%%%%%%%%%%%%%%%%%%
%  Method                                                                      %
%%%%%%%%%%%%%%%%%%%%%%%%%%%%%%%%%%%%%%%%%%%%%%%%%%%%%%%%%%%%%%%%%%%%%%%%%%%%%%%%
\section{Pseudo-Verlet Lists} \label{sec:pVl}

For the cell-based neighbour search, the simulation domain is first decomposed into cells of edge length $C_l \geqslant h$, and the particles corresponding to each cell are stored together. When searching for neighbours spanning a pair of cells, it is sufficient to loop over the particles of both cells and check if all possible particle pairs are within range of each other, e.g.

\begin{lstlisting}[caption={Simple iteration over neighbours.}, label=code:naive_pair]
for (int i = 0; i < count_a; i++)
  for (int j = 0; j < count_b; j++)
    r = particle_dist(parts_a[i], parts_b[j]);
    if (r < h)
      // Compute interaction.
\end{lstlisting}

\noindent where {\tt parts\_a} and {\tt parts\_b} are the particles in the neighbouring cells, and {\tt count\_a} and {\tt count\_b} their respective counts. Although much more efficient than the naive $\mathcal{O}(N^2)$ algorithm, less than $16\%$ of the particle pairs inspected will actually be within $h$ of each other (see Figure~\ref{fig:sph}), resulting in a large number of spurious pairwise distance calculations.

The pseudo-Verlet list algorithm \cite{ref:PG_Pseudo_Verlet_list} improves on the cell list algorithm by first projecting the particle positions from both cells onto the axis joining the cell centers. The particles in each cell are then sorted with respect to their position along this axis, resulting in the arrays {\tt dist} and {\tt index} containing the distance on the axis and its corresponding particle respectively. Given these two arrays for each cell, we can find the neighbours as follows

\begin{lstlisting}[caption={Pseudo-Verlet list iteration over neighbours.}, label=code:scalar_verlet_list]
for (int i = 0; i < count_a; i++)
  for (int j = 0; j < count_b @\colorbox{graybox}{\&\& dist\_b[j] < dist\_a[i] + h}@; j++)
    r = particle_dist(parts_a[index_a[i]], parts_b[index_b[j]]);
    if (r < h)
      // Compute interaction.
\end{lstlisting}

The {\tt dist} and {\tt index} arrays need to be pre-computed for every configuration of neighbouring cells, e.g. cell pairs sharing a common face, a common edge, or a common corner, which reduces to 13 distinct directions if symmetries are exploited. If in the inner loop we replace {\tt dist\_a[i] + h} with {\tt dist\_a[i] + h + max\_dx}, where {\tt max\_dx} is the maximum displacement of any particle in cell $a$, then we can re-use the sorted indices over several time steps, as is described in \cite{ref:PG_Pseudo_Verlet_list}.

Using this scheme, about 68\% of particle pairs inspected will be within range of each other, which is a significant improvement over the 16\% in the naive algorithm. The increased performance, however, comes at a cost in complexity: whereas the naive algorithm vectorises trivially, the pseudo-Verlet algorithm, with its additional loop exit condition and out-of-order access to the particle data, does not.

\begin{figure}
  \centering
  \begin{subfigure}[b]{0.55\textwidth}
    \includegraphics[width=1\linewidth]{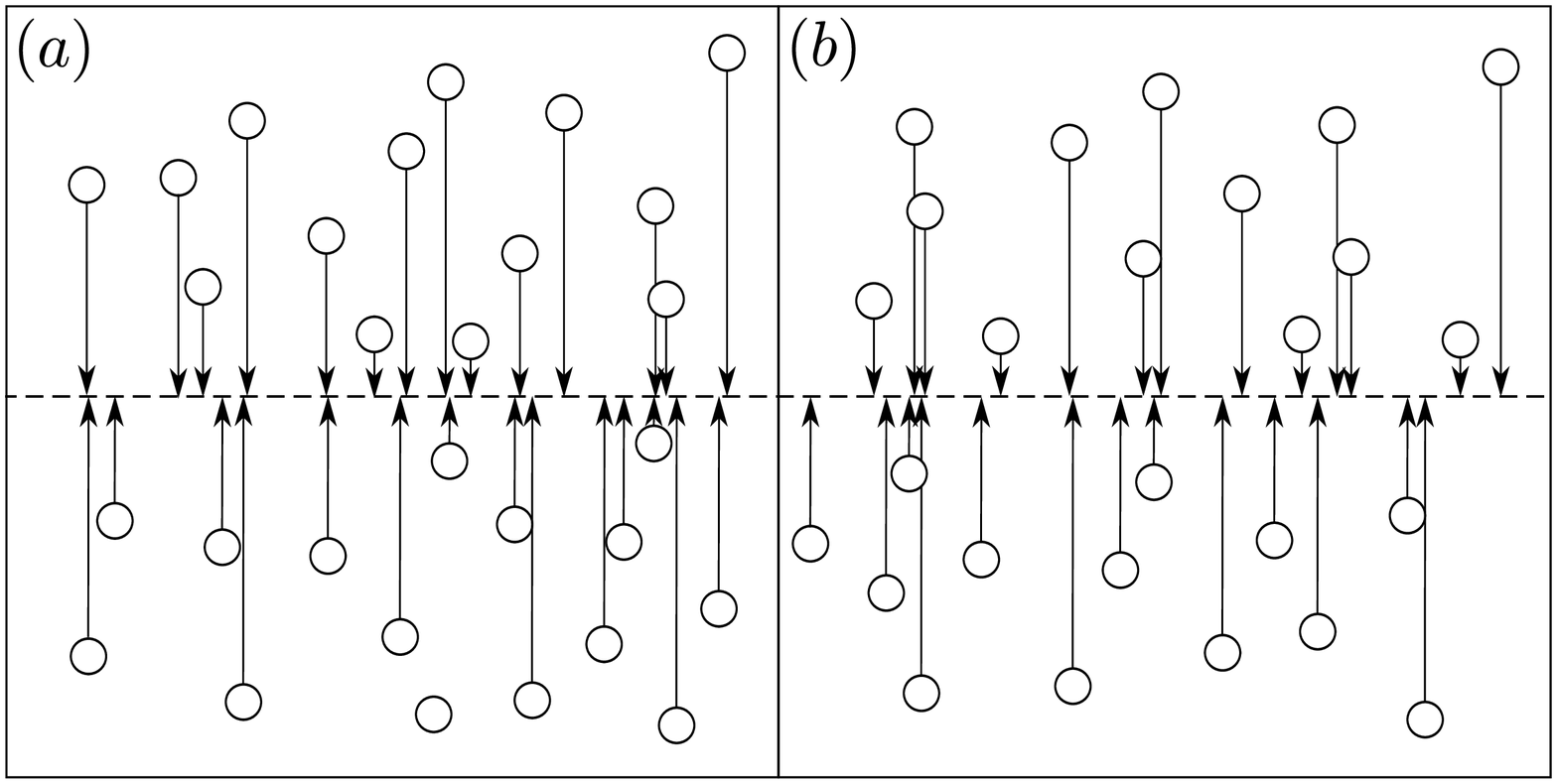}
  \end{subfigure}

  \begin{subfigure}[b]{0.55\textwidth}
    \includegraphics[width=1\linewidth]{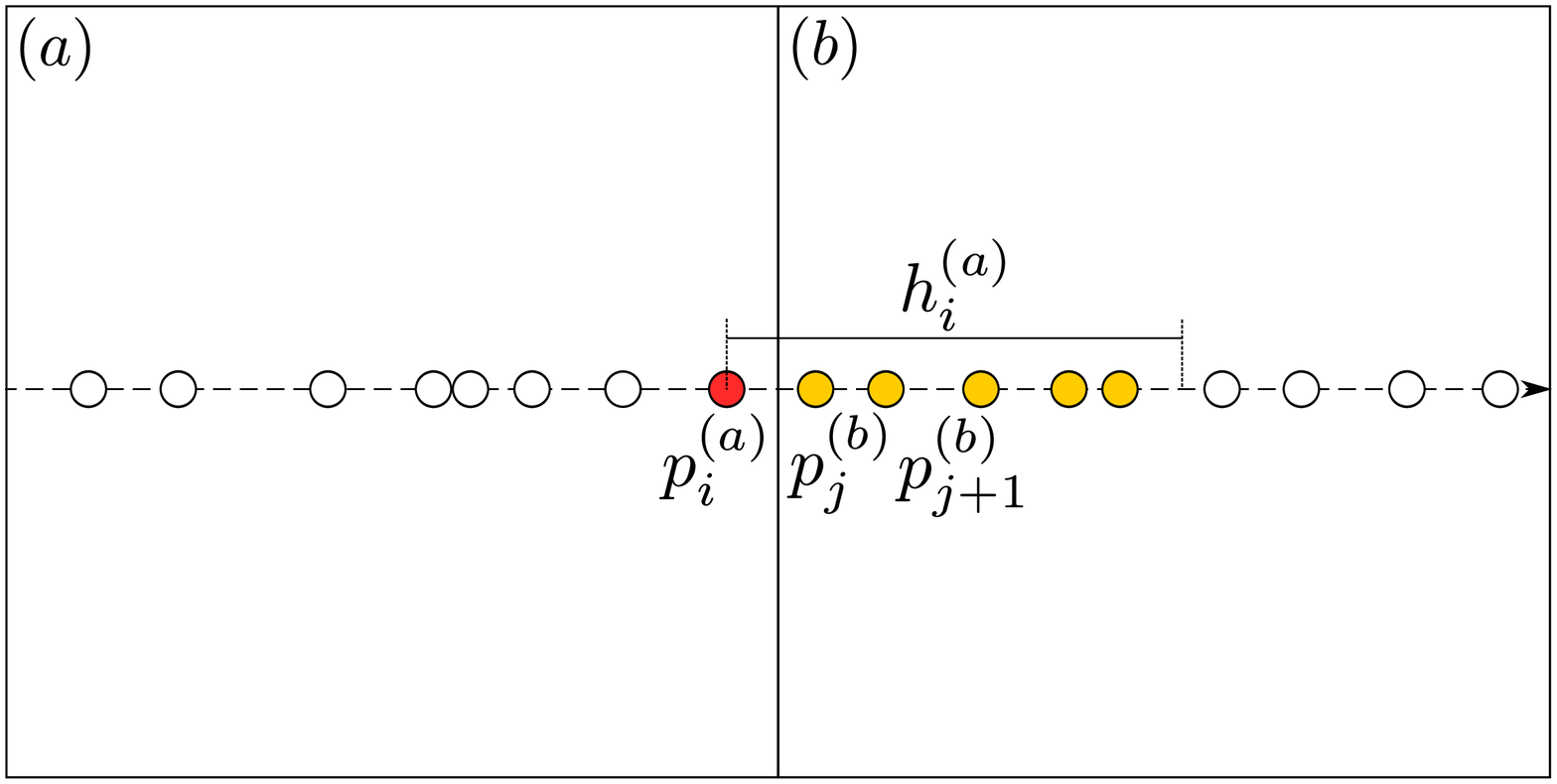}
  \end{subfigure}

  \caption{Performing cell-pair interactions using a sorted pseudo-Verlet list. \textit{Top:} Particles are projected onto the cell-pair axis and sorted according to their distance along the axis. The particles $p_i^{(a)}$ on the left (cell ($a$)) receive contributions from the particles $p_j^{(b)}$ on the right (cell ($b$)). \textit{Bottom:} Each particle is traversed and a pairwise distance calculation is only performed for particles in ($b$) that are within $h_i$ on the axis as this provides an upper-bound on the 3D distance. Interactions are later computed for the particles $p_i^{(a)}$ that obey $ \|p^{(a)}_i - p^{(b)}_j\|_2 < h_i^{(a)}$.}
  \label{fig:sort_parts}
\end{figure}

%%%%%%%%%%%%%%%%%%%%%%%%%%%%%%%%%%%%%%%%%%%%%%%%%%%%%%%%%%%%%%%%%%%%%%%%%%%%%%%%
%  Implementation                                                              %
%%%%%%%%%%%%%%%%%%%%%%%%%%%%%%%%%%%%%%%%%%%%%%%%%%%%%%%%%%%%%%%%%%%%%%%%%%%%%%%%
\section{SIMD Implementation}
\label{sec:simd_implementation}

The optimisations are split between the neighbour search part of the algorithm, which can be applied to any particle-based code, and the particle interaction function. 

Most compilers (Intel, GNU and Clang) fail to auto-vectorise the algorithm efficiently due to multiple exit conditions and conditional statements (see lines 2 and 4 of Code \ref{code:scalar_verlet_list}). We therefore decided to perform explicit vectorisation of the code using vector intrinsics\cite{ref:intel_intrinsics_guide}, which allows us to achieve the same level of performance across different compilers. In the following we outline a series of modifications that improve the performance of the code. 

\subsection{General SIMD Strategy}

The overall vectorisation strategy involved picking one particle, $p^{(a)}_i$, from the left-hand cell ($a$) and interacting it with a vector length of, $p^{(b)}_j$, particles that are within range from the right-hand cell ($b$). The particles, $p^{(a)}_i$, are traversed in reverse along the axis, i.e. starting from the cell interface and moving away from it. A pairwise distance calculation is only performed with particles, $p^{(b)}_j$ that lie within $h^{(a)}_i$, the cut-off radius of $p^{(a)}_i$, on the sorted axis. If any particles $p^{(b)}_j$ satisfy the criterion $\|p^{(a)}_i - p^{(b)}_j\|_2 < h^{(a)}_i$, where $\|\cdot\|_2$ denotes the Euclidean distance between two particles, an interaction is computed. The contributions from the $p^{(b)}_j$ that are not within range are masked out. This method is illustrated on Figure~\ref{fig:sort_parts}.

As the particle interaction functions are straight-forward, their vectorisation is not particularly interesting and we will therefore focus on the neighbour search part of the algorithm in the following.

\subsection{Particle Caching}
\label{sec:particle_caching}

We create a {\em particle cache} that only contains the subset of particle properties needed for an interaction. The particles are cached in sorted order along the cell-pair axis, using the sorted list of particle indices from the Verlet list. This cache is laid out using a structure of arrays (SoA) data layout and is only performed for sections of the code that include particle interaction. Although creating and filling this cache incurs an additional computational cost, this overhead is amortised with the improved memory access patterns for the SIMD implementation.

A side benefit is that the particle positions can be reduced from double to single precision values in this step. In order to retain numerical precision, we first shift the positions by the center of the cell-pair $c$, wrapping for boundary conditions, and then convert them to single precision, e.g. $x_{\mathtt{single}} = x_{\mathtt{double}} - c$. This allows for a more efficient use of the vector processing units whilst preventing numerical cancellation when computing the pairwise distances.

\subsection{Lowering the Number of Pairwise Distance Calculations}
\label{sec:max_d}

Computing the pairwise distances between particles (line 3 of Code \ref{code:scalar_verlet_list}) is the most time-consuming part of the neighbour search algorithm. In order to reduce the number of distance calculations, we find the {\em leftmost} particle in cell ($a$) which is in range of any of the particles in cell ($b$), e.g.

\begin{lstlisting}
int first_a = count_a;
while (first_a > 0 && dist_a[first_a - 1] + h_max_a > dist_b[0])
  first_a--;
\end{lstlisting}

We then introduce the array {\tt max\_index\_a} which contains, for each particle in cell ($a$), the index of the farthest possible interacting particle in cell ($b$). This array is constructed as follows:

\begin{lstlisting}
int temp = 0;
while (temp < count_b - 1 && 
      (dist_a[first_a] + parts_a[index_a[first_a]].h > dist_b[temp])) temp++;
max_index_a[first_a] = temp;
for (i = first_a + 1; i < count_a; i++)
  temp = max_index_a[i - 1];
  while (temp < count_b - 1 && 
        (dist_a[i] + parts_a[index_a[i]].h > dist_b[temp])) temp++;
  max_index_a[i] = temp;
\end{lstlisting}

\begin{figure}
  \centering
  \includegraphics[width=0.6\textwidth]{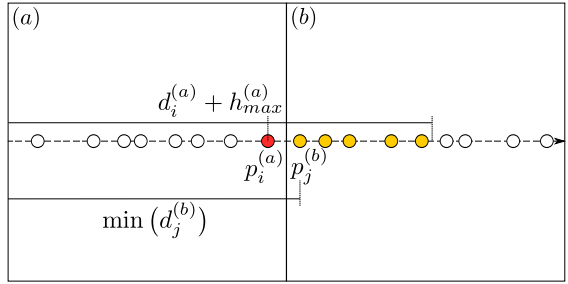}
  \caption{For a given particle $p_i^{(a)}$ (red circles), only the particles within a distance $d_i^{(a)}+h_{\rm max}^{(a)}$ (yellow circles) are candidates for interactions in cell ($b$). These particles are located between index {\tt 0} and {\tt max\_index\_a[i]} in the sorted array of particles of cell ($b$).}
   \label{fig:read_cache_subset}
\end{figure}

\noindent For each {\tt i}th particle in cell ($a$) from {\tt first\_a} to {\tt count\_a}, we only need to inspect the particles in cell ($b$) from {\tt 0} to {\tt max\_index\_a[i]} (see Figure~\ref{fig:read_cache_subset}). By consequence, only the particles from cell ($b$) from {\tt 0} to {\tt max\_index\_a[count\_a - 1]} need to be loaded into the particle cache. Note that we compute {\tt max\_index\_a} as an approximate value using only a single pass over the particles; we could compute an exact value but it would require $\mathcal{O}(N^2)$ operations.

Since we are computing interactions symmetrically, we compute the same indices for particles in cell ($b$) interacting with particles in cell ($a$), i.e. {\tt last\_b} and {\tt min\_index\_b} respectively, and cache the particles accordingly. We therefore have to be careful to cache the union of the set of particles from both cells required for interactions in both directions.

\subsection{Final Interaction Loop}
\label{sec:final_loop}

Once we have calculated the loop bounds, we find the particle neighbours and perform the interaction:

\begin{lstlisting}
for (int i = first_a; i < count_a; i++)
  for (int j = 0; j <= max_index_a[i]; j += SIMD_LENGTH)
    simd_r = particle_dist_simd(i, j);
    mask = (simd_r < SIMD(parts_a[index_a[i]].h));
    if (ANY(mask))
      particle_interact_simd(i, j, mask);
\end{lstlisting}

\noindent where {\tt particle\_dist\_simd(i, j)} returns a SIMD vector containing the pairwise distances between the cached particle data from cell ($a$) at position {\tt i} and the {\tt SIMD\_LENGTH} cached particle positions from cell ($b$) starting at index {\tt j}. Similarly, \texttt{particle\_interact\_simd(i, j, mask)} computes the interactions between the particle at index {\tt i} and the particles as of index {\tt j} using the provided {\tt mask}. Note that in practice, {\tt max\_index\_a[i]} is padded to a multiple of the SIMD width of the instruction set used. The results of the interactions on particle {\tt i} are stored in a SIMD vector and only aggregated horizontally and stored only once at the end of the innermost loop (not shown in code).~\\

%%%%%%%%%%%%%%%%%%%%%%%%%%%%%%%%%%%%%%%%%%%%%%%%%%%%%%%%%%%%%%%%%%%%%%%%%%%%%%%%
%  Results                                                                     %
%%%%%%%%%%%%%%%%%%%%%%%%%%%%%%%%%%%%%%%%%%%%%%%%%%%%%%%%%%%%%%%%%%%%%%%%%%%%%%%%
\section{Results}

In this section we present results obtained on different instruction sets using a standardized benchmark of the core compute kernels of the cosmological SPH code \swift\cite{ref:SWIFT_PASC}. The benchmark, named {\em test27cells}, computes the interactions between one cell containing $216$ randomly placed particles and its $26$ neighbours which are similarly populated. This allows us to probe the speed-up from vectorisation in all the relevant geometrical configurations, and is representative of the workload in actual production runs using \swift. All results reported here are based on revision \texttt{11518ff7} of \swift, which implements the algorithm described in Section~\ref{sec:simd_implementation}.

Internally, \swift uses a set of macros that map a common set of intrinsics to their equivalents for each specific instruction set. The same user code is used for all cases but the actual underlying compiled code will be architecture-specific. We currently support the AVX, AVX2 and AVX-512 architectures, but the common intrinsics can easily be extended to other instruction sets such as AltiVec or ARM NEON.

\subsection{Platforms, Compiler, and Methodology}

Table~\ref{table:machines_benchmark} lists the main characteristics of the machines used for our benchmarks. On all three systems we used the Intel compiler v.17.0.2 with the \texttt{-O3} flag alongside the respective SIMD-specific flags. We additionally used the \texttt{-no-vec} and \texttt{-no-simd} flags to produce the scalar code used in the comparisons.

\begin{table}[h]
  \centering
      \caption{Machines used for benchmarking. The KNL processor was placed in \textit{Flat-Quadrant} mode.}
    \label{table:machines_benchmark}
  \begin{tabular}{ clccc } 
    \hline
    \hline
    Machine Name & \makecell{Processor} & Cores & Vector ISA & Clock Rate [GHz]\\ 
    \hline
    COSMA-5 & Intel Xeon E5-2670 (Sandy Bridge) & 2 $\times$ 8 & AVX & 2.6 \\ 
    Hamilton & Intel Xeon E5-2650 (Broadwell) & 2 $\times$ 12 & AVX2 & 2.2 \\
    Kyll & Intel Xeon Phi 7210 (Knights Landing) & 1 $\times$ 64 & AVX-512 & 1.3 \\
    \hline
    \hline
  \end{tabular}
\end{table}

To obtain precise execution times we use the {\tt RDTSC} cycle counter and convert the cycle counts to milliseconds using the clock-speed of each platform. All benchmarks are run using a single thread and we report the median time of 5 independent runs.

\subsection{Idealised Particle Interaction Speed-ups}

We start by reporting the speed-up of the raw particle interactions, i.e. in the absence of any neighbour search. This highly idealised test assumes that all particles are always in range of each other, hence using SIMD vector instructions with full masks. This represents the hypothetical maximal speed-up that can be achieved for this problem and provides an upper-bound on the possible speed-up that can be achieved with the pseudo-Verlet list algorithm. This test interacts one particle with $2560$ other particles by directly calling the interaction function {\tt particle\_interact\_simd(i, j, mask)} with a full {\tt mask}. Table~\ref{table:raw_interaction_results} lists the times of both the scalar and vectorised functions.

\begin{table}[h]
  \centering
   \caption{Median times and corresponding vectorisation speed-ups for the idealized case of one particle directly interacting with $2560$ other particles without any distance checks.}
  \label{table:raw_interaction_results}
  \begin{tabular}{ clccc } 
    \hline
    \hline
    \makecell{Machine Name\\~} & \makecell{CFLAGS\\~} & \makecell{Scalar Time [ms] \\ (\texttt{-no-vec -no-simd})} & \makecell{Vectorised Time [ms] \\ ~} & \makecell{Speed-up\\~} \\ 
    \hline
    COSMA-5 & \texttt{-xAVX}          & $0.048$ & $0.0084$ & 5.66x \\
    Hamilton & \texttt{-xCORE-AVX2}    & $0.038$ & $0.0055$ & 6.77x \\ 
    Kyll & \texttt{-xMIC-AVX512}   & $0.170$ & $0.0079$ & 21.30x  \\ 
    \hline
    \hline
  \end{tabular}
\end{table}

The difference between the AVX and AVX2 results (both using 256-bit long vectors) is due to the use of Fused Multiply-Add {\em FMA} instructions in the AVX2 set. Similarly, the speed-up for the AVX-512 instruction set exceeds the vector length of 16 due to the use of FMAs.
These results demonstrate that, as expected due to their simple structure (no branching), the interaction functions themselves vectorise extremely well.

The results in Table~\ref{table:raw_interaction_results} compare favourably to those of \cite{ref:Luigi_Gadget} who reported a vectorisation speed-up of 6.62x on KNL (AVX-512) and 2.20x on Ivy-Bridge (AVX) when analysing a similar loop over neighbours in their improved version of the \gadget code\cite{ref:Gadget}. The difference might, in part, be explained by their use of double-precision arithmetic throughout.

\subsection{Pseudo-Verlet List SIMD Speed-ups}

As the number of particle pairs probed in the pseudo-Verlet algorithm depend on the orientation of the pair of cells, we break down the results into cell pairs sharing either a {\em corner}, an {\em edge}, or a {\em face}. We then compute a weighted total (i.e. $8\times\mathrm{corner}+12\times\mathrm{edge}+6\times\mathrm{face}$) to obtain the speed-up corresponding to a realistic scenario where all 26 cell-pair orientations have to be computed. Table~\ref{table:cell_orientation_results} lists the results obtained using the AVX instruction set on COSMA-5.

\begin{table}[h]
  \caption{Median times and corresponding speed-ups of the full SIMD pseudo-Verlet list algorithm implemented using the AVX instruction set on COSMA-5. Results for each geometrical configuration of the cell-pairs are reported as well as for the weighted total.}
  \label{table:cell_orientation_results}
  \centering
  \begin{tabular}{ cccc } 
    \hline
    \hline
    \makecell{Cell-pair \\ Orientation} & \makecell{Scalar Time [ms] \\ (\texttt{-no-vec -no-simd})} & \makecell{Vectorised Time [ms] \\ (AVX)} & \makecell{Speed-up\\~} \\ 
    \hline
    Corner 	    & $0.00035$ & $0.00070$ & 0.49x   \\
    Edge    	& $0.0052$ & $0.0035$ & 1.48x   \\ 
    Face  		& $0.082$ & $0.034$ & 2.41x  \\ 
    \hline
    {\em Total} & $0.56$ & $0.25$ & 2.21x  \\
    \hline
    \hline
  \end{tabular}
\end{table}

As expected, the results show a strong dependence of both the times and the speed-ups on the cell-pair orientation. Since a corner cell-pair will typically interact only one or two particles, the near-constant cost of populating the cache and computing the loop exit condition cannot be amortized by the higher speed at which the interactions can be processed. However, the total time spent in corner configurations is negligible and does not significantly affect the total speed-up. In \swift we therefore use the scalar code for the corner configurations, which leads to a weighted total speed-up over the 26 cell-pair configurations of 2.24x. 

Both the edge and face cases show speed-ups from vectorisation and the gain is larger when more particles interact (face vs. edge) as the constant vectorisation cost can be more easily amortized. The best speed-up obtained (2.41x) is far from the theoretical maximum for AVX (8x) but compares well to the raw interaction speed-ups (5.66x). The difference in performance arises from the constant cost of populating the cache, and additional complexity of computing distances and masking of out-of-range interactions which leads to partially full masks, especially for the edge-orientation case. Given the complexity of the algorithm we conclude that 2.41x is a good speed-up.

In Table~\ref{table:pseudo_verlet_list_simd_results} we present the median times and speed-ups for the weighted sum of all cell-pair orientations, for the three available instructions sets.

\begin{table}[h]
  \caption{Median times and corresponding speed-ups of the full SIMD pseudo-Verlet list algorithm for the weighted sum of all cell-pair orientations implemented using the AVX, AVX2 and AVX-512 instruction sets.}
  \label{table:pseudo_verlet_list_simd_results}
  \centering
  \begin{tabular}{ clccc } 
    \hline
    \hline
    \makecell{Machine Name\\~} & \makecell{CFLAGS\\~} & \makecell{Scalar Time [ms] \\ (\texttt{-no-vec -no-simd})} & \makecell{Vectorised Time [ms]\\~} & \makecell{Speed-up\\~} \\ 
    \hline
    COSMA-5 & \texttt{-xAVX}        & $0.56$ & $0.25$ & 2.24x \\
    Hamilton & \texttt{-xCORE-AVX2}  & $0.49$ & $0.20$ & 2.43x \\ 
    Kyll & \texttt{-xMIC-AVX512} & $1.98$ & $0.49$ & 4.07x \\ 
    \hline
    \hline
  \end{tabular}
\end{table}

The newer and extended instruction sets show larger speed-ups compared to AVX. This is the result of:

\begin{itemize}
	\item The raw interactions being faster (Table~\ref{table:raw_interaction_results}),
    \item The dedicated masking instructions,
    \item The FMAs that enter the distance calculation and, 
    \item In the case of AVX-512, the larger vector lanes.
\end{itemize}

We finally note that a naive implementation of a double {\tt for} loop over all particles (See Section~\ref{sec:pVl}), interacting all particles within range without using any clever algorithms, runs in 24.49ms  on the COSMA-5 system. Even if this implementation were to be sped-up using SIMD instructions by the ideal factor of 8 for AVX, this code would still be slower than our pseudo-Verlet list implementation by more than a factor of 12x. This demonstrates the importance of using clever algorithms, and investing the effort in modifying them for improved SIMD vectorisation.

%%%%%%%%%%%%%%%%%%%%%%%%%%%%%%%%%%%%%%%%%%%%%%%%%%%%%%%%%%%%%%%%%%%%%%%%%%%%%%%%
%  Conclusions                                                                 %
%%%%%%%%%%%%%%%%%%%%%%%%%%%%%%%%%%%%%%%%%%%%%%%%%%%%%%%%%%%%%%%%%%%%%%%%%%%%%%%%
\section{Conclusions}

We presented an efficient SIMD implementation of the pseudo-Verlet list algorithm commonly used in particle-based codes to interact particles located in neighbouring cells. A particle cache in sorted order alongside an accurate estimation of the upper bounds of the loop trip count were used to reduce the number of distance calculations to perform and ensure optimal data alignments. An accumulator vector was used to temporarily store the results before pushing back to memory once all neighbour interactions of a given particle were performed. 

When implemented in the \swift code using the AVX, AVX2 and AVX-512 instruction sets this algorithm reached speed-ups of 2.24x, 2.43x and 4.07x respectively when compared to a scalar version. These results demonstrate the importance of, not just hand-coding vectorised assembly loops, but also of developing better algorithms alongside vectorisation to achieve maximal performance.

%%%%%%%%%%%%%%%%%%%%%%%%%%%%%%%%%%%%%%%%%%%%%%%%%%%%%%%%%%%%%%%%%%%%%%%%%%%%%%%%
%  Acknowledgement                                                             %
%%%%%%%%%%%%%%%%%%%%%%%%%%%%%%%%%%%%%%%%%%%%%%%%%%%%%%%%%%%%%%%%%%%%%%%%%%%%%%%%
\section{Acknowledgements}
This work would not have been possible without Lydia Heck's help and expertise. We thank the \swift team for their help and input on this project as well as John Pennycook and Georg Zitzelberger from {\sc Intel} for their help with details of the intrinsics. This work is supported by {\sc Intel} through establishment of the ICC as an {\sc Intel} parallel computing centre (IPCC). This work used the DiRAC Data Centric system at Durham University, operated by the Institute for Computational Cosmology on behalf of the STFC DiRAC HPC Facility (\url{www.dirac.ac.uk}). This equipment was funded by BIS National E-infrastructure capital grant ST/K00042X/1, STFC capital grants ST/H008519/1 and ST/K00087X/1, STFC DiRAC Operations grant ST/K003267/1 and Durham University. DiRAC is part of the National E-Infrastructure. This work made use of the facilities of the Hamilton HPC Service of Durham University.

%%%%%%%%%%%%%%%%%%%%%%%%%%%%%%%%%%%%%%%%%%%%%%%%%%%%%%%%%%%%%%%%%%%%%%%%%%%%%%%%
%  References                                                                  %
%%%%%%%%%%%%%%%%%%%%%%%%%%%%%%%%%%%%%%%%%%%%%%%%%%%%%%%%%%%%%%%%%%%%%%%%%%%%%%%%
\bibliographystyle{ieeetr}
\bibliography{references}

\end{document}